\documentclass[aps,pra,reprint,groupedaddress,superscriptaddress,nofootinbib]{revtex4-2}
\usepackage{graphicx,amsmath,amssymb,amsbsy,subfigure,hyperref,bbm,times,txfonts,float}
\usepackage{subfigure,hyperref,bbm,times}
\usepackage[T1]{fontenc}
\usepackage{braket}
\usepackage{epsfig}
\usepackage{color}
\usepackage{graphicx}
\usepackage{dcolumn}
\usepackage{bm}

\providecommand{\openone}{\leavevmode\hbox{\small1\kern-3.8pt\normalsize1}}

\usepackage{soul}

\hypersetup{
	colorlinks=true,
	linkcolor=blue,
}
\graphicspath{{figure/}}

\begin{document}
	
	\preprint{PRA/123-QED}
	
	\title{Monitoring variations  of refractive index via Hilbert-Schmidt speed and applying this phenomenon to improve quantum metrology}

	\author{Seyed Mohammad Hosseiny}
	
	\affiliation{Physics Department, Faculty of Sciences, Urmia University,P.B. 165, Urmia, Iran}

	\author{Hossein Rangani Jahromi}
	\email{h.ranganijahromi@jahromu.ac.ir}
	\affiliation{
		Physics Department, Faculty of Sciences, Jahrom University, P.B. 74135111, Jahrom, Iran}

	\author{Mahdi Amniat-Talab}
	\affiliation{Physics Department, Faculty of Sciences, Urmia University,P.B. 165, Urmia, Iran}

	\date{\today}
	
	\begin{abstract}
		Effective nonlinear optical interactions are essential for many applications in modern photonics. In this paper, we investigate the role of the nonlinear response of a material to improve quantum metrology. In particular, the collective optical behavior of an atomic ensemble is applied to enhance frequency estimation through one of the atoms. Moreover, we introduce Hilbert-Schmidt speed, an easily computable theoretical tool, to monitor the variations of linear as well as nonlinear refractive indices and evaluate the strength of the nonlinear response of optical materials. Furthermore, we illustrate that quantum Fisher information and Hilbert-Schmidt speed can efficiently detect negative permittivity and refractive index, which is of great importance from a practical point of view.
	\end{abstract}
	
	\maketitle

	\section{Introduction \label{introduction}}
\par 
Nonlinear optics \cite{boyd2020nonlinear} has been a rapidly developing field of science in recent years. It is based on the phenomena connected to the interaction of intense laser fields with matter. In fact, nonlinear optics focuses on the interactions of light with matter under conditions in which the nonlinear response of the atoms plays a significant role. These phenomena cover a broad range of applications \cite{garmire2013nonlinear}, including spectroscopy \cite{mukamel1999principles,backus2021probing}, telecommunications \cite{schneider2004nonlinear},  optical data storage as well as processing \cite{cotter1999nonlinear,glezer1996three,wu2020high},  and quantum information technologies \cite{leach2010quantum,howell2004realization,zhang2021high}, especially quantum metrology \cite{alodjants2022enhanced}.
	\par 

One of the important parameters of interest to estimate in linear as well as nonlinear optics is the unknown frequency of a laser beam or radiation emitted in a spectroscopy experiment. Moreover, evaluating and controlling frequency or wavelength are vital in
different situations to protect body organs or avoid damage to the materials. For example,  light at a wavelength sufficiently long, not damaging biological materials, can be applied to gain a resolution requiring normally a much shorter wavelength \cite{boyd2020nonlinear}. Moreover, the importance of the frequency dependence of the breakdown fields in different materials has been investigated \cite{soileau1978frequency,wood2003laser,buscher1965frequency,apostolova2022femtosecond,weyl2020physics}. In addition,  the choice of frequency has a profound effect on the design of linear colliders \cite{braun2003frequency}. Furthermore, the frequency dependence of the refractive index of various materials, called dispersion, is a well-known phenomenon in linear and nonlinear optics. Additionally, a phase match based frequency estimation can be used to enhance the ranging precision of linear frequency modulated continuous wave radars \cite{shen2015phase}.
More interestingly, the capability of decreasing the quantum noise in gravitational wave interferometers considerably depends on the frequency which should be detected \cite{danilishin2012quantum}. Moreover, quantum noise in the audio-band spectrum can be reduced by frequency-dependent squeezing schemes \cite{polino2020photonic,brown2017broadband}.
These arguments motivate us to investigate frequency estimation and particularly explore how optical tools can be applied to achieve optimal measurement.

\par 
The field of quantum metrology \cite{helstrom1969quantum,paris2009quantum,giovannetti2006quantum, giovannetti2011advances,degen2017quantum,pirandola2018advances,pezze2018quantum,huang2019cryptographic,liu2022optimal,jahromi2021hilbert,yang2022variational,colombo2022time,garbe2020critical,salvatori2014quantum,barbieri2022optical,fallani2022learning,chu2022thermodynamic,rangani2019weak}  has provided valuable tools, such as the  the classical Fisher information  and the quantum Fisher information (QFI)  \cite{petz2011introduction,seveso2019discontinuity,genoni2011optical,zanardi2008quantum,gacon2021simultaneous,vsafranek2018simple,jafarzadeh2020effects,demkowicz2020multi,jahromi2019quantum},  for estimating the sensitivity of measurement devices and unknown parameters. Such analyses are usually focused on finding optimal measurement strategies, with respect to environmental parameters as well as initial conditions, to achieve measurement sensitivities better than the standard quantum limit \cite{kura2020standard}.
However, achieving the optimal strategies for a measurement performed on a single atom with respect to the collective behavior of the atomic ensemble is rarely investigated. 
\par
In this paper, we demonstrate how the optical behavior of an atomic ensemble can be used to enhance the frequency estimation of input laser fields. This approach is different from the recent works which have focused on this quantum frequency estimation \cite{huelga1997improvement,escher2011general,demkowicz2012elusive,albarelli2020quantum,haase2016precision,wineland1992spin,wineland1994squeezed,seveso2020quantum,leibfried2004toward,dorfman2016nonlinear,udem2002optical,naghiloo2017achieving,giovannetti2011advances}. Moreover, we propose  Hilbert-Schmidt speed \cite{gessner2018statistical,rangani2022searching}, which is a special type of quantum statistical speed, to monitor the variations of linear and nonlinear refractive indices of an optical material. In particular, we also propose quantum Fisher information and Hilbert-Schmidt speed as efficient tools to characterize materials exhibiting negative permittivity as well as refractive index \cite{smith2004metamaterials}, important active fields of research in modern Optics.

\par

The concept of materials with negative refraction was implied by Veselago in 1968 \cite{veselago1967electrodynamics}. This kind of material has negative magnetic permeability ($ \mu<0 $) and negative dielectric permittivity ($\epsilon<0 $). These materials, possessing several interesting properties, such as a reverse Doppler shift and reverse Cherenkov radiation,  allow novel applications like subwavelength imaging and remarkable control over light propagation. In addition to double-negative materials, a material for which only one of the material parameters (either magnetic permeability or electric permittivity) has a negative value is also of research interest and may be applied to realize left-handedness \cite{fredkin2002effectively,alu2003pairing,castles2020active,zhang2006producing,hossain2021modified,liu2011metamaterials}. In this paper, the efficiency of quantum Fisher information and Hilbert-Schmidt speed to detect these phenomena is addressed. Moreover, we introduce Hilbert-Schmidt speed as an easily computable theoretical tool to predict the variations of linear as well as nonlinear refractive indices of optical materials. We know that the \textit{phase matching }condition of nonlinear optics sensitively depends on the variation of the refractive index within the sample being imaged \cite{muller19983d}.
It is well known that under this condition, the output wave maintains a fixed phase relation versus the nonlinear polarization and can extract energy most efficiently from the input waves. We also show that monitoring the refractive index can be applied to improve the frequency estimation.

The paper is structured as follows: In Sec. \ref{Preliminaries}, the concepts of quantum Fisher information and Hilbert-Schmidt speed are explained. Then, after introducing two models in Secs. \ref{FirstModel} and \ref{Second Model}, we 
discuss how those concepts can be related to optical responses of the material.
Finally,  Sec. \ref{Conclusion} is devoted to summarizing and discussing the most important results.

\section{Preliminaries}\label{Preliminaries}
\subsection{Quantum Fisher information}  
The fundamental question when investigating the sensitivity of a quantum state with respect to other parameters is the following: By performing measurements on similar systems affected by some unknown parameter $\eta$ (which may be, for example, a parameter quantifying the magnitude of a gravitational field or rotation, acceleration), how accurately can $\eta$ be estimated? The answer is expressed by the quantum Cramer-Rao bound \cite{braunstein1994statistical}, dictating that the smallest resolvable change in the parameter $\eta$ is given by	
		\begin{equation}
		\delta \eta=\left(1 / \sqrt{F_{\eta}}\right)
	\end{equation}
	where $F_{\eta}$ denotes the quantum Fisher information (QFI), which for pure states can be simply written as
	\begin{equation}
		F_{\eta}=4\left[\langle\dot{\psi} \mid \dot{\psi}\rangle-|\langle\psi \mid \dot{\psi}\rangle|^{2}\right],
	\end{equation}
	where $|\dot{\psi}\rangle=(\partial / \partial \eta)|\psi\rangle$. According to the theory of quantum estimation, an increase in QFI indicates the improvement of the optimal accuracy of estimation. Therefore, QFI is an efficient quantity measuring the maximum information on parameter $ \eta $, extractable from a given measurement process.
	
		\subsection{Hilbert-Schmidt speed}
	
	Considering quantum state $ \rho(\eta)$, one can define the   Hilbert–Schmidt speed (HSS) as \cite{gessner2018statistical,jahromi2020witnessing}
	\begin{align}\label{HSSS}
		HSS_{{\eta}}(\rho\big)=\sqrt{\frac{1}{2}\text{Tr}\bigg[\bigg(\dfrac{d\rho(\eta)}{d\eta}\bigg)^2\bigg]},
	\end{align}
	which is an important quantifier of quantum statistical speed with respect to parameter $\eta$.
	It is interesting to note that no diagonalization of  $ \text{d}\rho(\varphi)/\text{d}\eta $ is required for HSS computation.
	\par
	The HSS is known as an efficient tool for detecting non-Markovianity as well as improving quantum phase estimation in $ n $-qubit open quantum systems.
	Generally speaking, because both HSS and QFI  are quantum statistical speeds corresponding, respectively,  to the \textit{Hilbert–Schmidt} and  \textit{Bures distances} \cite{ozawa2000entanglement,braunstein1994statistical}, it is fascinating to investigate in detail how they can be
	related to each other in different scenarios.
	\par 
	Because the explicit expressions for the HSS and QFI in our models have  cumbersome forms, they are  not reported here. 
	\section{Model I (four-level system) }\label{FirstModel}
		\begin{figure}
		\centering
		\includegraphics[width=0.5\linewidth]{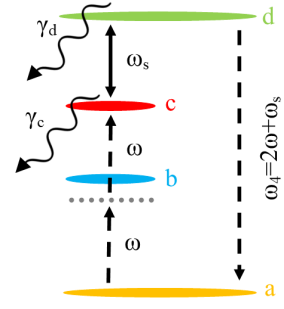}
		\caption{A typical situation, consisting of four-level atoms, for observing electromagnetically induced transparency..}
		\label{figmodel1}
	\end{figure}
	
	We focus on a setup in which strong and weak laser fields  $ E_{s}  $ and $ E $ at frequencies, respectively, $\omega_{s}$ and $\omega$ are applied to a four-level atomic system in order to generate radiation at the sum frequency $\omega_{4}=2 \omega+\omega_{s}$ (see Fig. \ref{figmodel1}) \cite{boyd2020nonlinear}. It is known that by allowing the field at frequency $\omega_{s}$ to be a strong saturating field one, we can efficiently eliminate linear absorption at the $a \rightarrow d$ transition frequency while gaining a large four-wave-mixing susceptibility. This kind of absorption elimination is called electromagnetically induced transparency (EIT). We assume that the conditions necessary to achieve the EIT are satisfied. 
	It is emphasized that the nonlinear response can remain considerably large even when linear absorption at the output frequency vanishes in the EIT technique.

	Our first goal is to investigate how the sum-frequency generation process is related to estimating frequencies of the driving laser fields and the quantum statistical speed. In particular, we study the nonlinear response leading to sum-frequency generation by analyzing its behavior via quantum Fisher information and Hilbert-Schmidt speed computed for the frequencies of the driving fields.
	\par
	The Hamiltonian of the atomic system can be split into two terms as
	$ H = H_{0} + V(t)$
	in which $ H_{0}  $ denotes the Hamiltonian of the free atom. Moreover, using the rotating-wave and electric-dipole approximations, one can see that $ V(t) $, designating the interaction energy of the atom with the externally driving radiation fields, can be expressed as
	\begin{equation}\label{Vmodel1}
		V(t)=-\mu\left(E e^{-i \omega t}+E_{s}^{*} e^{i \omega_{s} t}\right),
	\end{equation}
	where  $\mu  $ represents the atomic dipole moment. Moreover, the Rabi frequencies $ \Omega_{b a}  $, $ \Omega_{cb}  $, and $ \Omega_{dc}  $ are defined by
	
	\begin{equation}
		\begin{aligned}
			&V_{ba}=-\mu_{ba}E e^{-i \omega t}=-\hbar \Omega_{b a} e^{-i \omega t} ,\\
			&V_{cb}=-\mu_{cb} E e^{-i \omega t}=-\hbar \Omega_{c b} e^{-i \omega t}, \\
			&V_{dc}=-\mu_{dc} E_{s} e^{-i \omega_{s} t}=-\hbar \Omega_{d c} e^{-i \omega_{s} t},
		\end{aligned}
	\end{equation}
	where the notations $ \bra{i} V \ket{j} \equiv V_{ij}$ and $ \bra{i} \mu \ket{j} \equiv \mu_{ij}$ are adopted.
	In addition,  the detuning factors 
	\begin{equation}
		\delta_{1}=\omega-\omega_{b a}, \quad \delta_{2}=2 \omega-\omega_{c a} \quad \text { and } \quad \Delta=\omega_{s}-\omega_{d c},
	\end{equation}
	in which $ \omega_{b a}= \omega_{b }-\omega_{a}$, $ \omega_{ca}= \omega_{c }-\omega_{a}$, and $ \omega_{dc}= \omega_{d }-\omega_{c}$ denote the transition frequencies, are introduced.
	It should be noted that the energies are measured  relative to that of the ground state $ \ket{a} $.

	We should compute the wavefunction to find the response leading to sum-frequency generation.  In the interaction picture, it can be expressed as \cite{boyd2020nonlinear}
	\begin{equation}
		\begin{aligned}
			\ket{\psi(t)}=& C_{a}(t) \ket{a}+C_{b}(t)  \ket{b} e^{-i \omega t} \\
			&+C_{c}(t)  \ket{c} e^{-i 2 \omega t}+C_{d}(t)  \ket{d} e^{-i\left(2 \omega+\omega_{c}\right) t},
		\end{aligned}
	\end{equation}
	obeying Schrödinger's equation in the form
	\begin{equation}\label{Shrodinger}
		i \hbar \frac{\partial \ket{\psi}}{\partial t}=H\ket{\psi} \quad \text { with } \quad H(t)=H_{0}+V(t)
	\end{equation}

	Solving the Schrödinger equation perturbatively in terms of  $\Omega_{b a}$ and $\Omega_{c b}$ but to all orders in $\Omega_{d c}$ and  taking $\dot{C}_{i}=0$ for the steady-state solution, we find that \cite{boyd2020nonlinear}
	
	\begin{equation}
		\begin{aligned}
			&C_{a}=1,C_{b}=-\Omega_{b a} / \delta_{1},-C_{c} =\frac{C_{b} \Omega_{c b}}{\delta_{2}}+\frac{C_{d} \Omega_{d c}^{*}}{\delta_{2}},&\\&C_{d}=-\frac{\Omega_{d c} \Omega_{c b} \Omega_{b a}}{\delta_{1}\left[\delta_{2}\left(\delta_{2}+\Delta\right)-\left|\Omega_{d c}\right|^{2}\right]}.
		\end{aligned}
	\end{equation}

	Now, calculating the induced dipole moment at the sum frequency, i.e., 
	$
	\tilde{p}=\langle\psi|\mu| \psi\rangle
	$
	one   finds that the third-order non-linear optical susceptibility is given by \cite{boyd2020nonlinear}
	$$
	\chi^{(3)}=\frac{-N \mu_{a d} \mu_{d c} \mu_{c b} \mu_{b a}}{3 \epsilon_{0} \hbar \delta_{1}\left[\delta_{2}\left(\delta_{2}+\Delta\right)-\left|\Omega_{d c}\right|^{2}\right]} ,
	$$
	where $ \epsilon_{0}  $ represents the vacuum permittivity and $ N $ denotes the number density of atoms. 
	The effects of damping to this result	 can be added  by replacing $\delta_{2}$ and  $\delta_{2}+\Delta$ with, respectively,  $\delta_{2}+i \gamma_{c}$ and  $\delta_{2}+\Delta+i \gamma_{d}$,  leading to  
	\begin{equation}
		\chi^{(3)}=\frac{-N \mu_{a d} \mu_{d c} \mu_{c b} \mu_{b a}}{3 \epsilon_{0} \hbar \delta_{1}\left[\left(\delta_{2}+i \gamma_{c}\right)\left(\delta_{2}+\Delta+i \gamma_{d}\right)-\left|\Omega_{d c}\right|^{2}\right]},
	\end{equation}
	where $\gamma_{d}$ and $\gamma_{c}$ represent the decay rates of the probability amplitudes to be in levels $d$ and $c$, respectively.

	\par 
	$| \chi^{(3)} |$ quantifies the strength of the third-order nonlinear optical response of the material driven by the laser field. 
	In the presence of   strong laser field $ E_{s} $, the refractive index,  experienced by a weak wave,  can be written as \cite{reshef2019nonlinear}
	$
	n=n_{0}+n_{2} I,
	$
	in which  $n_{0}$ denotes the usual (i.e., low-intensity or linear ) refractive index. Moreover, $ n_{2} $, called the nonlinear refractive index, characterizes the strength of the optical nonlinearity induced by the strong field. If the system exhibits negligible absorption, it is given by 
	$
	n_{2}=\frac{3}{4 n_{0}^{2} \epsilon_{0} c} \chi^{(3)},
	$
	in which $ c $ denotes the speed of light. In Ref. \cite{del2004relation} the general connection between $ n_{2} $ and $ \chi^{(3)} $ has been discussed.  In addition,  $I=2 n_{0} \epsilon_{0} c |E_{s}
	|^{2}$ represents the time-averaged intensity of the strong wave. Furthermore, the linear refractive index $ n_{0} $ is connected with the linear susceptibility $ \chi^{(1)}  $  and linear dielectric constant $ \epsilon^{(1)} $ via  
	\begin{equation}
		n_{0} =\sqrt{\epsilon^{(1)} }=\sqrt{1+\chi^{(1)} }. 
	\end{equation}
	
	\par
	The imaginary part of the refractive index characterizes the absorption of radiation which occurs due to the transfer of population from the atomic ground state to some excited state. Strictly speaking, the low-intensity  normal (i.e., linear)  \textit{absorption coefficient } can be represented in terms of the linear susceptibility $ \chi^{(1)} $ through
	\begin{equation}
		\alpha_{0}= \chi^{(1)''} \omega/c,
	\end{equation}
	where  the real and imaginary parts of the linear susceptibility have been defined as $\chi^{(1)} =\chi^{(1)'} +i\chi^{(1)''} $.  The absorption coefficient of many material systems decreases when measured using high laser intensity. Considering an incident laser radiation with intensity $ I $, one finds that the (high-intensity) absorption coefficient is given by $ \alpha(I)=\alpha_{0}/(1+I/I_{s}) $ in which $ I_{S} $ denotes  the saturation intensity for which $ \alpha(I_{S})=\alpha_{0}/2  $. However, in this paper, we investigate a weak laser field that is not strong enough to considerably correct its own absorption. Nevertheless,  the strong laser field $ E_{s} $, coupling the two upper energy levels, can affect the absorption of the weak field $ E $ through another mechanism called electromagnetically induced transparency.

		\subsection*{Relation between   strength of non-linear response and quantum statistical speeds }\label{RelationOmegas}
	In this section, we reveal important relationships between the non-linear response of the material to the optical driving fields and the quantum statistical speeds of a single four-level atom located inside the material.  Computing the quantum Fisher information (QFI) and Hilbert-Schmidt speed (HSS) versus frequency of the weak or strong field, we see that they exhibit completely similar qualitative behaviors. We also show that they can be used as efficient tools to measure the degree of the non-linear response of the material to the driving laser fields. 
	\subsubsection{Quantum statistical speeds with respect to $ \omega_{s} $ }\label{RelationOmegas1}
	
	\begin{figure}[ht]
		\subfigure[] {\includegraphics[width=7cm]{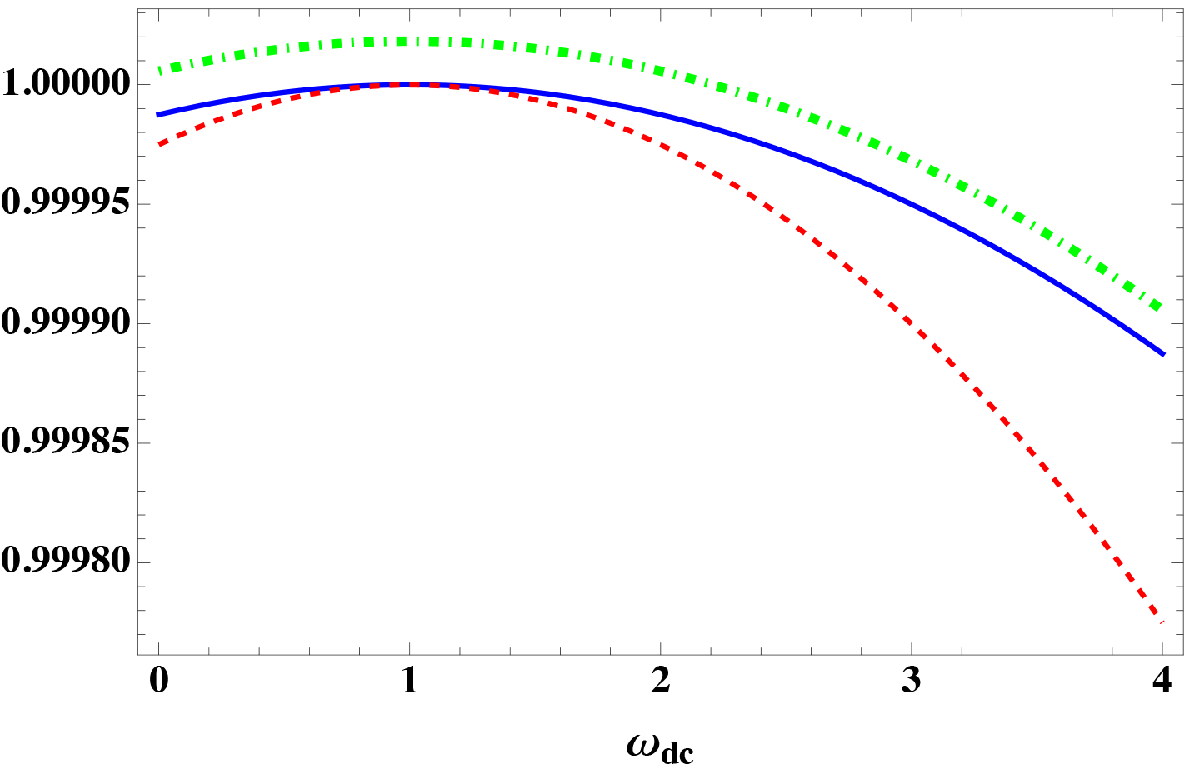}\label{QfiomegaS11}} 
		\hspace{0.5mm}
		\subfigure[] { \includegraphics[width=7cm]{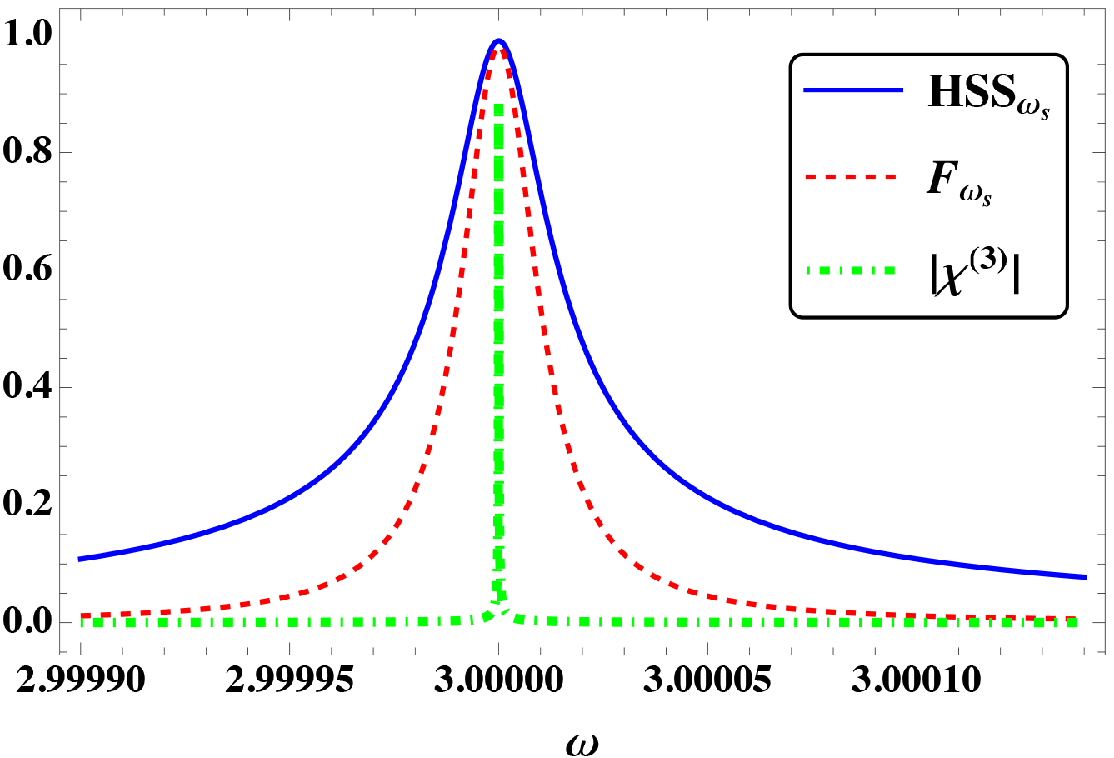}\label{QfiomegaS12}}
		\hspace{0.5mm}
		\subfigure[] { \includegraphics[width=7cm]{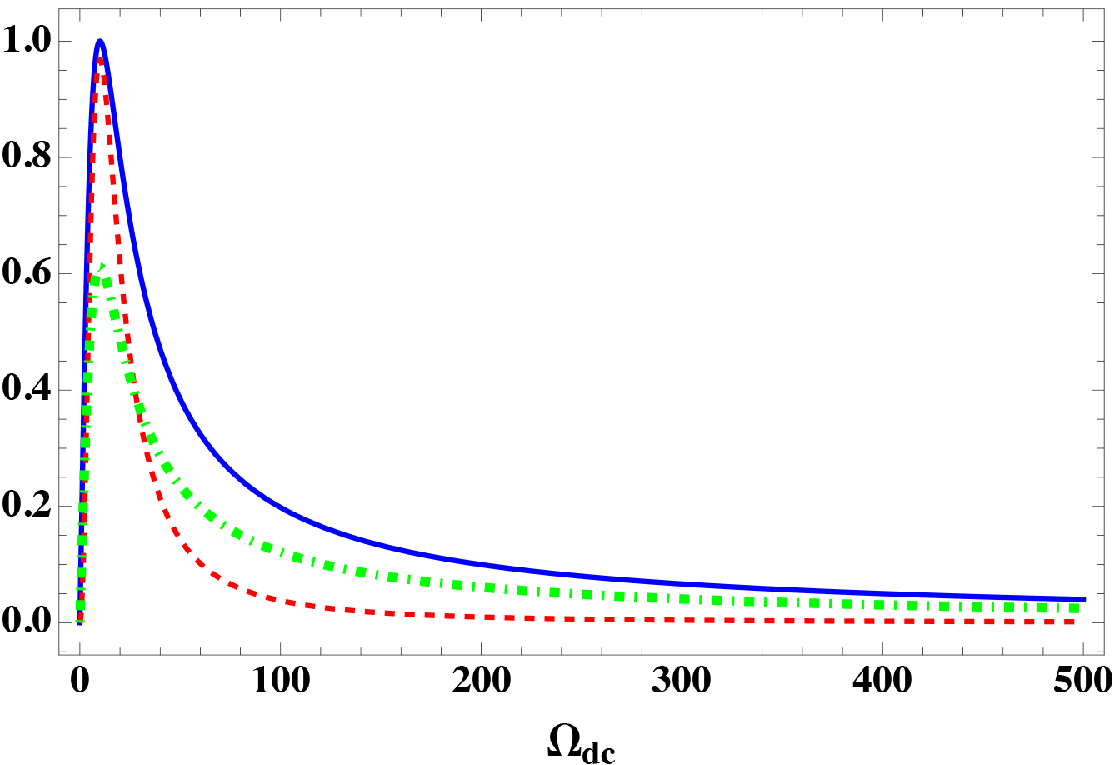}\label{QfiomegaS13}}
		\caption{Four-level system in the presence of damping: (a) Normalized quantum Fisher information $ F_{\omega_{s}} $, Hilbert-Schmidt speed $ HSS_{\omega_{s}} $, and  degree of non-linear response $ |\chi^{(3)} |$ as functions of $ \omega_{d c} $  for 
			$\Omega_{c b}=0.00001 ,  \Omega_{b a}= 0.000011,  \Omega_{dc}=10, \omega=3, \omega_{s}=1, \omega_{b a}=3.1, \omega_{c a}=6,  \gamma_{c}=1,  \gamma_{d}=100$; 
			(b)  The same quantities versus $ \omega $  for 
			$   \omega_{b a}=3,   \omega_{dc}=1.001$. (c) The same quantities versus $ \Omega_{dc} $  for 
			$  \omega=3, \omega_{s}=1.0001, \omega_{b a}=3.01, \omega_{c a}=6, \omega_{dc}=1.001.$}
		
		\label{QfiomegaS1}
	\end{figure}

		Figure \ref{QfiomegaS11}    compares the variations of the QFI, HSS, computed with respect to $ \omega $, and $ |\chi^{(3)} |$ versus $ \omega_{d c} $.  It is clear that if $\delta_{2} $=2$\omega $-$\omega_{ca} =0$, the best estimation occurs when $\Delta _{1}$=$\omega _{s}$-$\omega _{dc}$=0, i.e., $\omega _{dc}$=$\omega _{s}$. Therefore, the most precise frequency estimation of the strong field is obtained when it is resonant with the transition frequency $\omega _{dc}$. 
	
	Now, we investigate how control parameters  $ \omega $ and $ \Omega_{d c} $ should be fixed to achieve the optimal estimation of $ \omega_{s}  $.  Figure \ref{QfiomegaS12}  demonstrates that
	the maximum point of the QFI rises when $\delta _{1}$=$\omega $-$\omega _{ba}$=0, i.e., $\omega $=$\omega _{ba}$. Hence, the weak laser field should be tuned to resonance with the transition frequency $\omega _{ba}$ to achieve the most accurate estimation. 
	
	\par 
	In  Figs.  \ref{QfiomegaS11}, \ref{QfiomegaS12}, and \ref{QfiomegaS13}  we observe that the variations of the QFI and HSS, calculated with respect to $ \omega_{s} $,  are in perfect agreement with each other. In addition, $ HSS_{\omega_{s}}  (F_{\omega_{s}} ) $ can be used to predict the strength of the nonlinear response. In fact, 
	the point at which the HSS is maximized versus each of the parameters: ${\omega_{dc}, \omega, \Omega_{dc}} $ exactly reveals the value for which the strongest nonlinear response of the medium occurs. Moreover, because the maxima of the QFI and $   |\chi^{(3)} |$ coincide, the nonlinear response can be considered as a source for achieving the optimal frequency estimation of the strong field.

	\par
	It is also found that an increase in the Rabi frequency $ \Omega_{d c} $ first enhances the frequency estimation until the maximized nonlinear response of the medium is achieved (see Fig. \ref{QfiomegaS13}). Then, we see that an increase in the $ \Omega_{d c} $ results in suppression of the nonlinear response as well as the accuracy of the estimation. 
	
	\subsubsection{Quantum statistical speeds with respect to $ \omega $
	}\label{RelationOmega1}

	\begin{figure}[ht]
		\subfigure[] {\includegraphics[width=7cm]{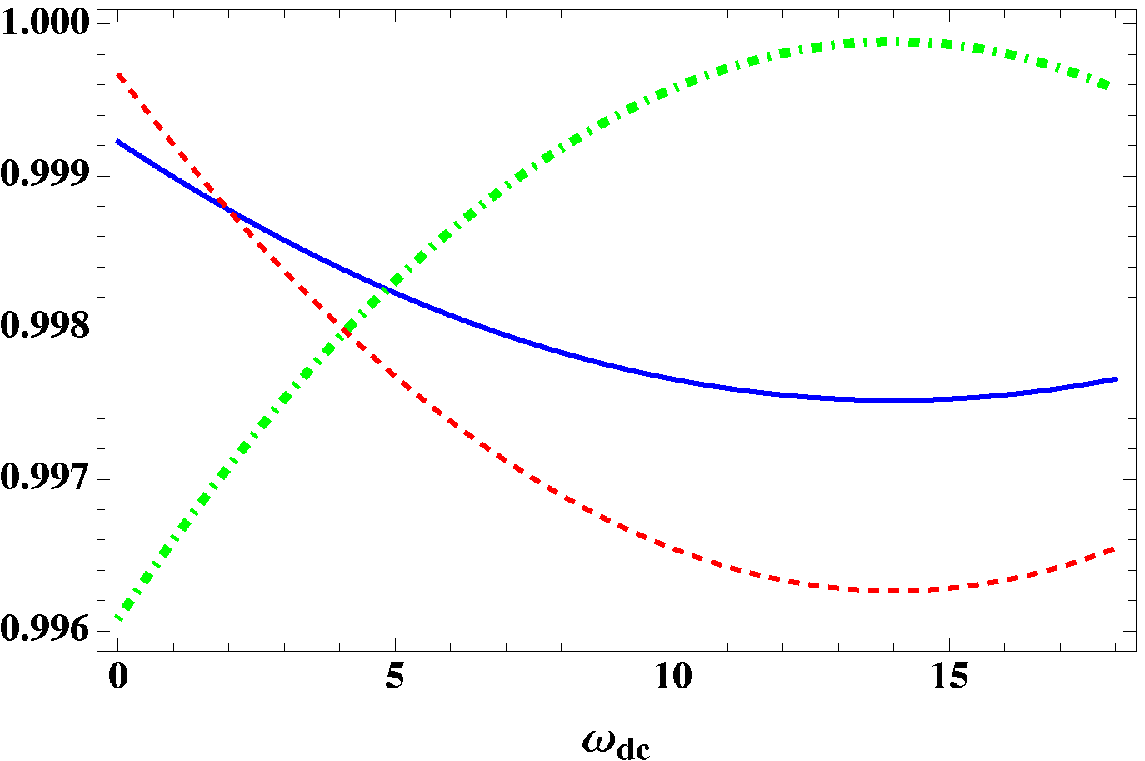}\label{Qfiomega11}} 
		\hspace{5mm}
		\subfigure[] { \includegraphics[width=7cm]{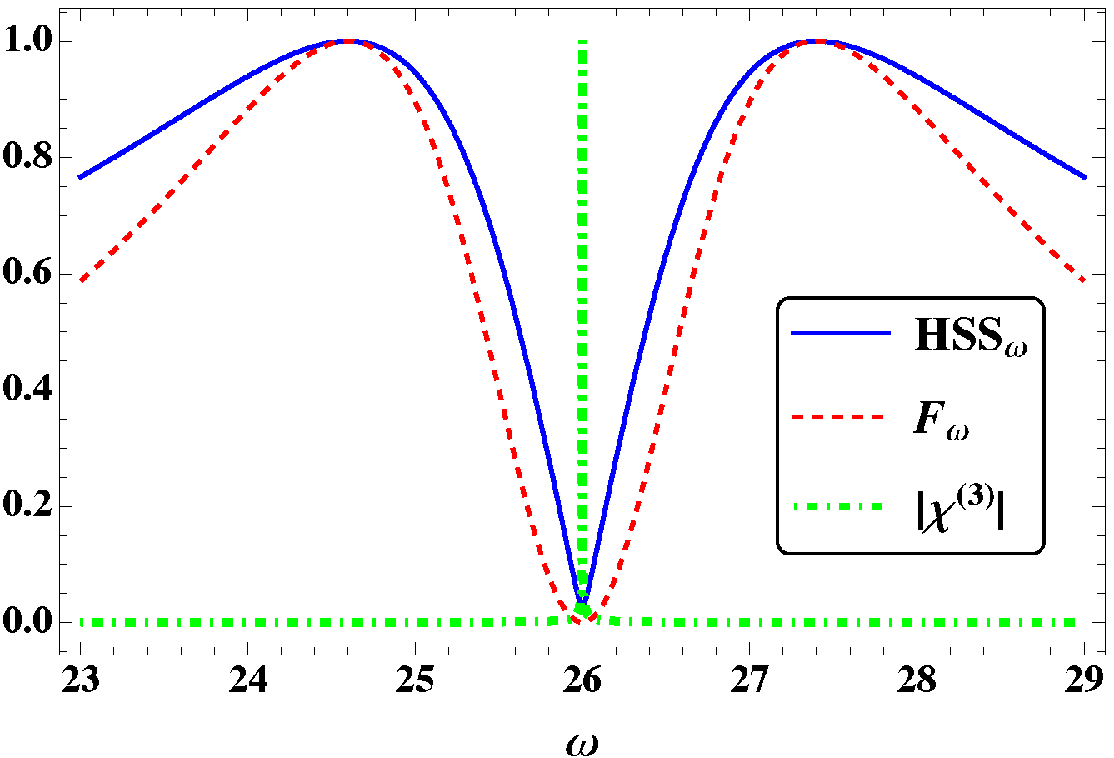}\label{Qfiomega12}}
		\caption{Four-level system in the presence of damping: (a) Comparison among normalized quantum Fisher information $ F_{\omega} $, Hilbert-Schmidt speed $ HSS_{\omega} $, and  degree of non-linear response $ |\chi^{(3)} |$ versus $ \omega_{d c} $  for 
			$\Omega_{c b}= 1 $, $ \Omega_{b a}= 1.4 $, $ \Omega_{dc}=100, \omega=26, \omega_{s}=14, \omega_{b a}=26.01, \omega_{c a}=52,  \gamma_{c}=100,  \gamma_{d}=60$; 
			(b)  The same quantities versus $ \omega $  for 
			$  \omega_{b a}=26,  \omega_{dc}=14.001,  \gamma_{c}=1,  \gamma_{d}=100$. }
		\label{Qfiomega1}
	\end{figure}

	Figure \ref{Qfiomega1} demonstrates that computation of the QFI and HSS with respect to $ \omega $ can provide us with valuable information on the nonlinear response of the material. 
	Investigating their variations versus $ \omega_{d c} $ or $ \omega $, we see their common minimum point exactly coincide with the maximum point of $   |\chi^{(3)} |$. Therefore, when the nonlinear response of the material is at the highest level versus $ \omega_{d c} $ or $ \omega $, the frequency estimation of the weak field is minimal while that of the strong field is maximal.
	Hence, we cannot simultaneously estimate the frequencies of the driving fields with the best accuracy.
	Again the HSS is introduced as an efficient figure of merit for revealing the strength of the nonlinear response in the medium.

	\section{Model II (three-level system)}\label{Second Model}
	\begin{figure}
		\centering
		\includegraphics[width=0.5\linewidth]{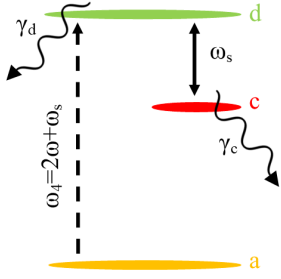}
		\caption{Another typical situation, consisting  of three-level atoms, for observing electromagnetically induced transparency..}
		\label{figmodel2}
	\end{figure}

	Now we introduce another model in which a nonlinear medium consisting of three-level atoms is driven by the radiation of amplitude $ E_{4} $  produced in the process of sum-frequency generation. The linear absorption at the frequency of this radiation can be essentially eliminated through the EIT technique. 
	In detail, as illustrated in Fig. \ref{figmodel2}, the linear absorption at frequency $\omega_{4}$ vanishes by an intense saturating field of amplitude $E_{s}$ at frequency $\omega_{s}$. 
	Including states $a, d$, and $c$ in the atomic wavefunction and working in the interaction picture,  we can write the wavefunction as
	\begin{equation}
		\ket{\psi(t)}=C_{a}(t) \ket{a}+C_{d}(t) \ket{d} e^{-i \omega_{4} t}+C_{c}(t) \ket{c}e^{-i\left(\omega_{4}-\omega_{s}\right) t},
	\end{equation}
	satisfying the Schrödinger's equation (\ref{Shrodinger}) with
	\begin{equation}
		V=-\mu\left(E_{4} e^{-i \omega_{4} t}+E_{s}^{*} e^{i \omega_{s} t}\right).
	\end{equation}
	Moreover, the Rabi frequencies $ (\Omega, \Omega_{s}) $ and the detuning factors  $ (\delta, \Delta) $ are introduced by the following equations
	\begin{equation}
		\begin{aligned}
			\bra{d}V\ket{a} &=-	\bra{d}\mu\ket{a} E_{4} e^{-i \omega_{4} t} =-\hbar \Omega e^{-i \omega_{4} t}, \\
			\bra{c}V\ket{d}&=-	\bra{c}\mu\ket{d} E_{s}^{*} e^{i \omega_{s} t}=-\hbar \Omega_{s}^{*} e^{i \omega_{s} t},
		\end{aligned}
	\end{equation}
	
	\begin{equation}
		\delta \equiv \omega_{4}-\omega_{d a} \quad \text { and } \quad \Delta \equiv \omega_{s}-\omega_{d c}.
	\end{equation}
	\par 
	Inserting $\ket{\psi(t)}$  into Schrödinger's equation (\ref{Shrodinger}) 
	and solving  the equations of motion for the coefficients $C_{j}$, one can determine  the evolved state of the  three-level atom.  We intend to solve the differential equations  perturbatively in terms of  $\Omega$  but to all orders in $\Omega_{s}$ and  take $\dot{C}_{i}=0$ for achieving the steady-state solution, leading to \cite{boyd2020nonlinear}
	
	\begin{equation}
		\begin{aligned}
			C_{a}=1,
			C_{c}=\dfrac{-\Omega-\delta C_{d}}{\Omega_{s}},
			C_{d}=\frac{\Omega(\delta-\Delta)}{\left|\Omega_{s}\right|^{2}-\delta(\delta-\Delta)}.
		\end{aligned}
	\end{equation}

	\par
	Calculating  the  induced  dipole moment $ \tilde{p}=\bra{\psi} \mu \ket{\psi}=p(\omega_{4})\text{e}^{-i\omega_{4}t}+c.c.$  and then the polarization $ P=N p \equiv \epsilon_{0} \chi^{(1)} E$ leads to the following expression for linear susceptibility \cite{boyd2020nonlinear}:
	\begin{equation}
		\chi^{(1)}=\frac{N\left|\mu_{d a}\right|^{2}}{\epsilon_{0} \hbar} \frac{(\delta-\Delta)}{\left|\Omega_{s}\right|^{2}-(\delta-\Delta) \delta}.
	\end{equation}
	The effects of damping can modelled by replacing $\delta$ by $\delta+i \gamma_{d}$ and $\Delta$ by $\Delta+i\left(\gamma_{c}-\gamma_{d}\right)$:
	\begin{equation}
		\chi^{(1)}=\frac{N}{\hbar} \frac{\left|\mu_{d a}\right|^{2}\left(\delta-\Delta+i \gamma_{c}\right)}{\left|\Omega_{s}\right|^{2}-\left(\delta+i \gamma_{d}\right)\left(\delta-\Delta+i \gamma_{c}\right)}.
	\end{equation}

	\subsection*{Relation between   strength of linear response and quantum statistical speeds }\label{Relation2}

	We aim to explore the relationship among the QFI, HSS, computed for the three-level atom, and the linear response of the medium, composed of a collection of the three-level atoms,  to the strong laser field. Again, we see that the qualitative behaviors of the HSS and QFI perfectly resemble. In addition, we introduce the linear response of the medium to the laser fields as a key tool to improve the frequency estimation.
	
	\subsubsection{Quantum statistical speeds with respect to $ \omega_{s} $ }\label{RelationOmegas2}
	\begin{figure}[ht]
		\subfigure[] {\includegraphics[width=7cm]{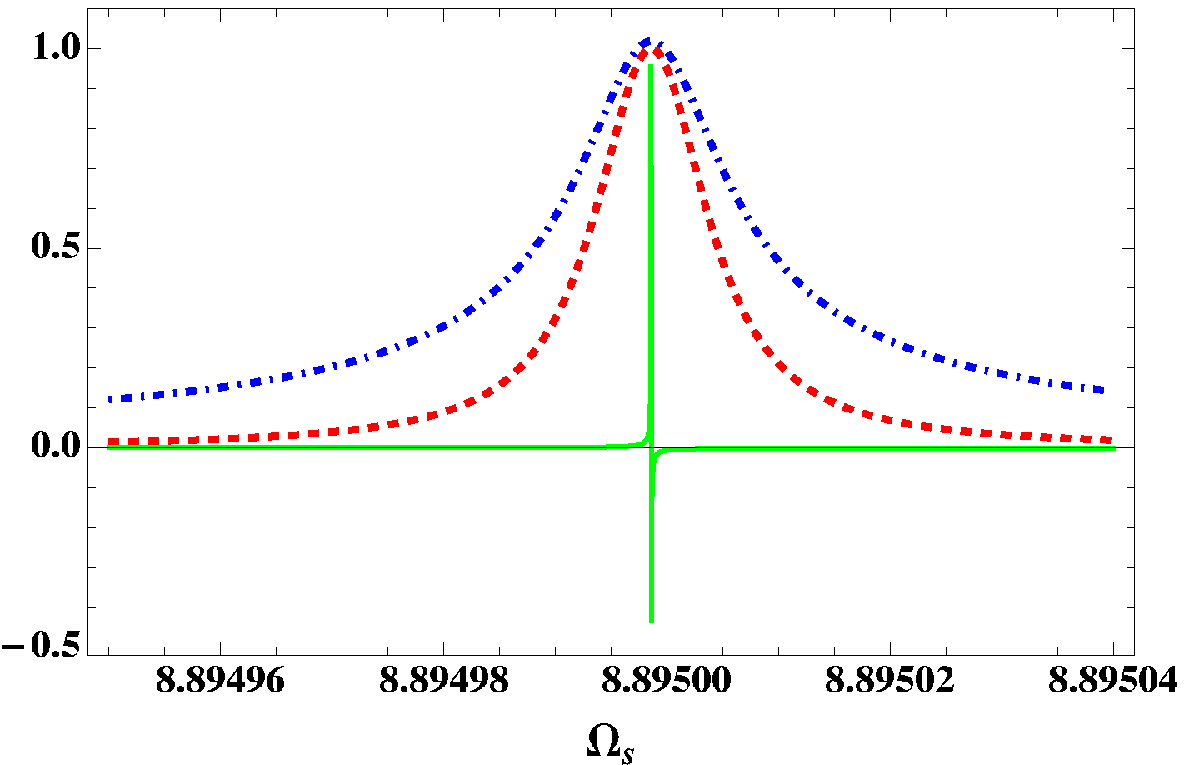}\label{OmegaS2a}} 
		\hspace{0.5mm}
		\subfigure[] { \includegraphics[width=7cm]{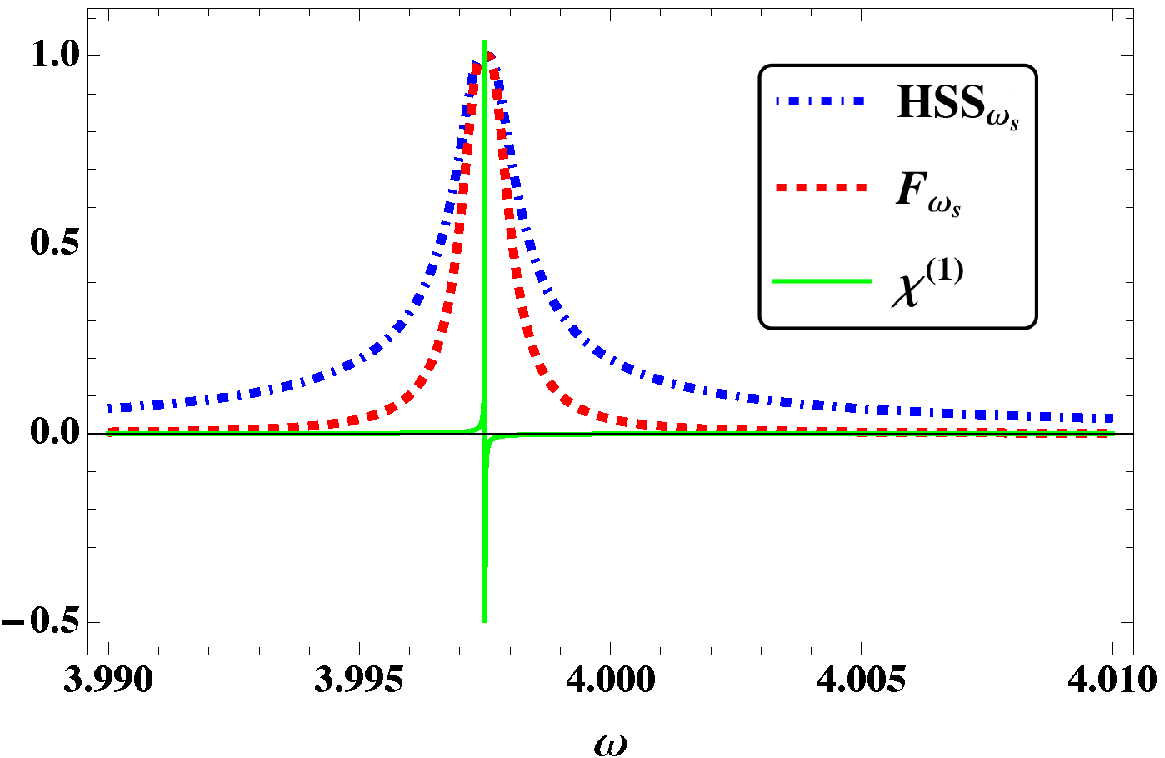}\label{OmegaS2b}}
		\hspace{0.5mm}
		\subfigure[] { \includegraphics[width=7cm]{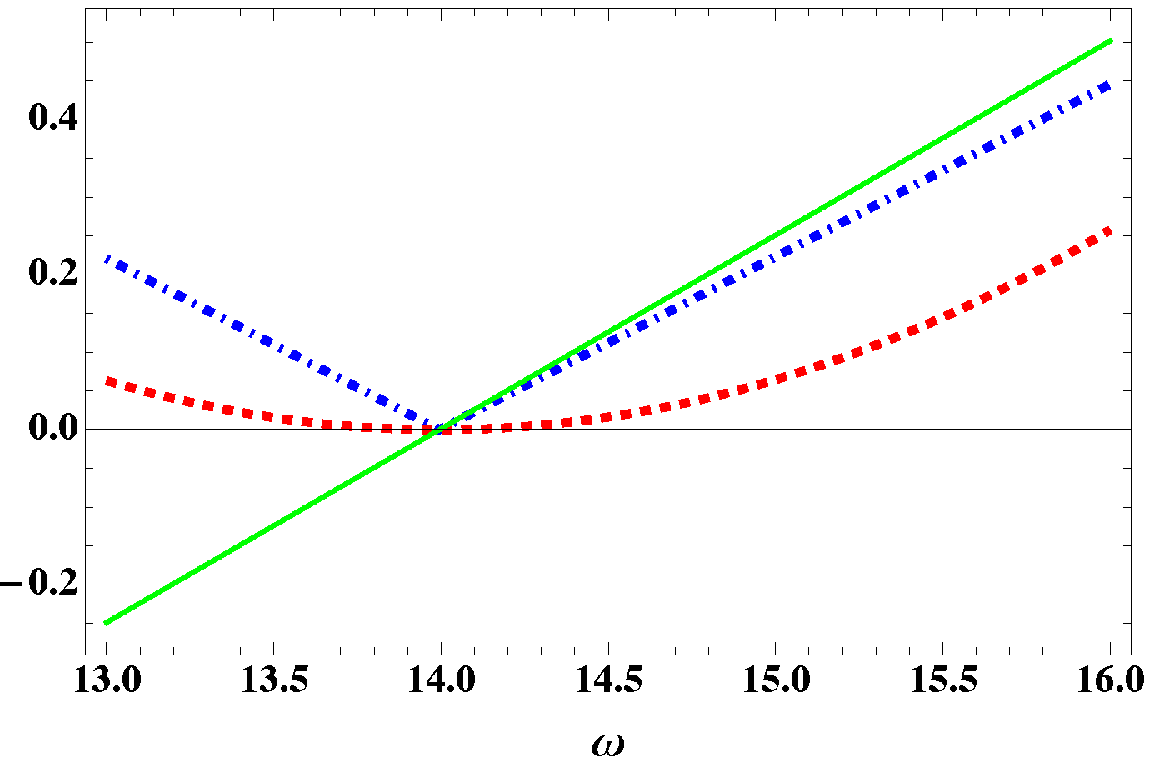}\label{OmegaS2c}}
		\caption{Three-level system in the absence of damping: (a) Normalized quantum Fisher information $ F_{\omega_{s}} $, Hilbert-Schmidt speed $ HSS_{\omega_{s}} $, and  linear susceptibility  $ \chi^{(1)} $ as functions of $ \Omega_{s} $  for 
			$\Omega=0.00001 ,  \omega_{da}=20, \omega=4.65, \omega_{dc}=1.8, \omega_{s}=1.81$; 
			(b)  The same quantities versus $ \omega $  for 
			$\Omega=0.001 ,  \Omega_{s}=10, \omega_{da}=20, \omega_{dc}=2, \omega_{s}=2.01$. (c)  The same quantities versus $ \omega $  for 
			$\Omega=0.00001 ,  \Omega_{s}=100, \omega_{da}=30, \omega_{dc}=2.01, \omega_{s}=2.$}
		\label{OmegaS2}
	\end{figure}
	
	\begin{figure}[ht]
		\subfigure[] {\includegraphics[width=5cm]{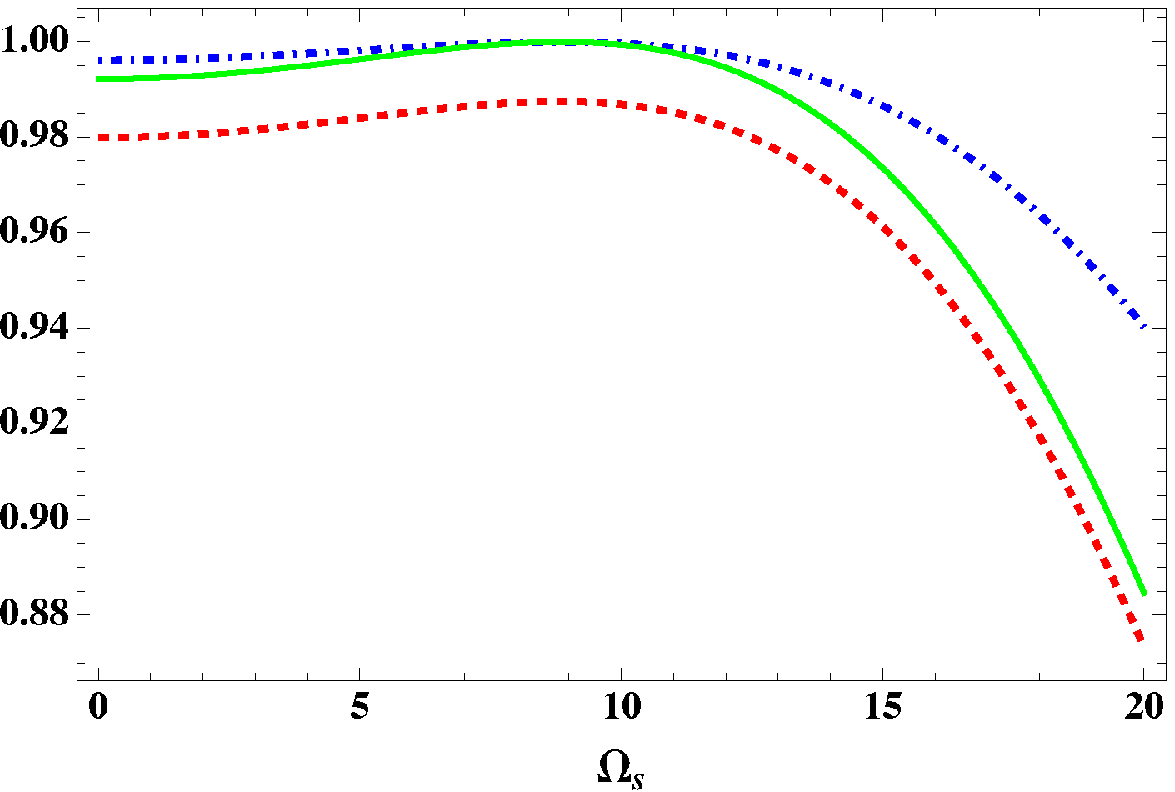}\label{OmegaS2deca}} 
		\hspace{0.5mm}
		\subfigure[] { \includegraphics[width=5cm]{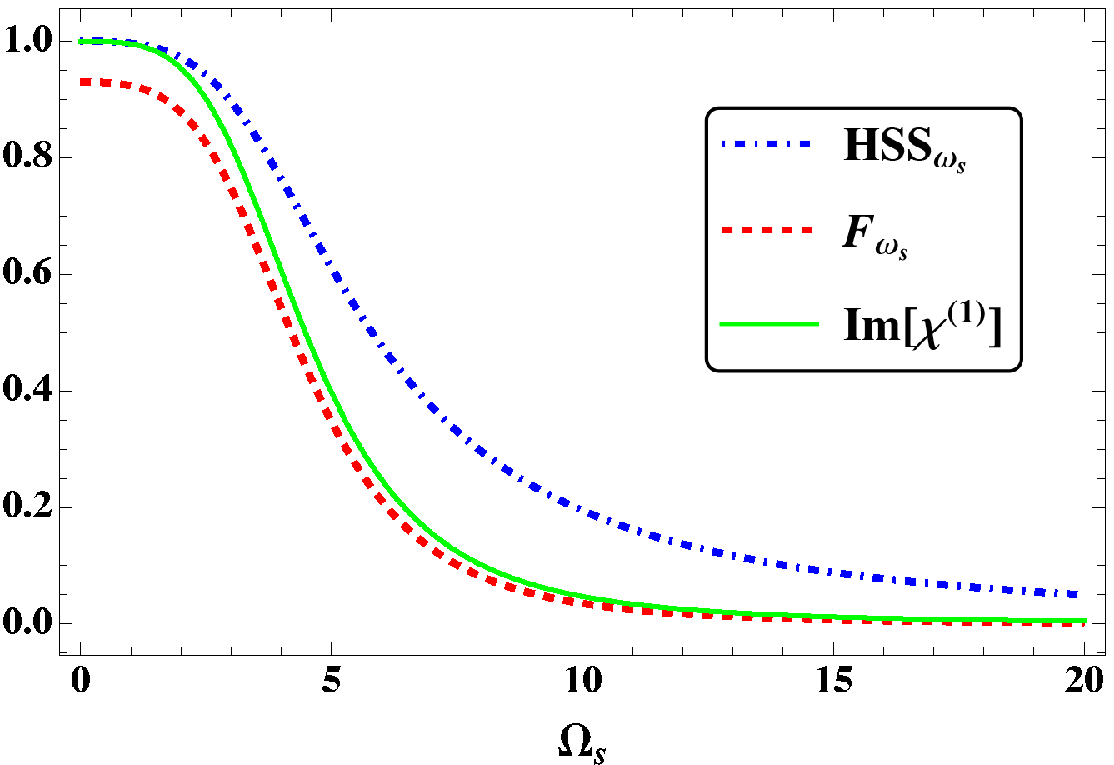}\label{OmegaS2decb}}
		\hspace{0.5mm}
		\subfigure[] { \includegraphics[width=5cm]{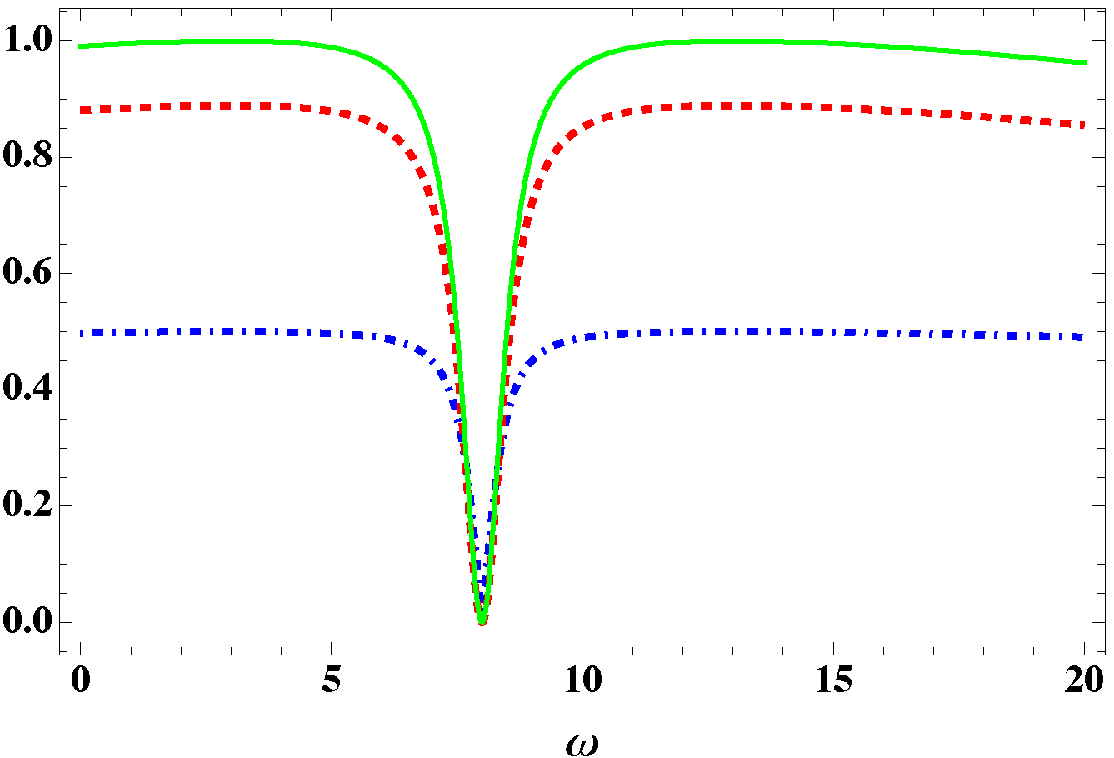}\label{OmegaS2decc}}
		\hspace{0.5mm}
		\subfigure[] { \includegraphics[width=5cm]{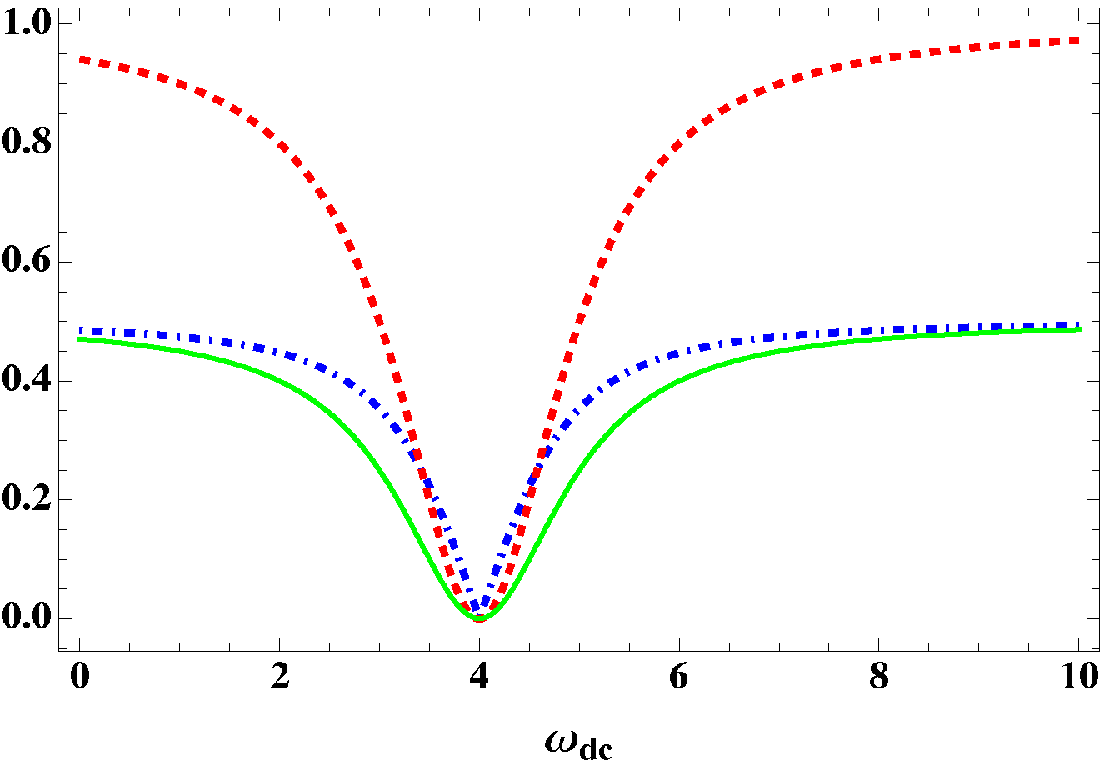}\label{OmegaS2decd}}
		\caption{Three-level system in the presence of damping: (a) Comparison among normalized quantum Fisher information $ F_{\omega_{s}} $, Hilbert-Schmidt speed $ HSS_{\omega_{s}} $, and the   imaginary part of the linear susceptibility  $ \text{Im}[\chi^{(1)}] $  versus $ \Omega_{s} $  for 
			$\Omega= 0.00001$, $  \omega_{da}=20, \omega=4.65, \omega_{dc}=1.8, \omega_{s}=1.81,  \gamma_{c}=0.01,  \gamma_{d}=100$; 
			(b)  The same quantities versus $ \Omega_{s} $  for 
			$  \omega=9$;  (c) The same quantities versus  $ \omega $  for 
			$ \Omega_{s}=10,  \omega_{dc}=4, \omega_{s}=4.001,  \gamma_{c}=0.001$;   (d) The same quantities versus  $ \omega_{dc} $  for 
			$ \Omega_{s}=10,  \omega=8, \omega_{s}=4,  \gamma_{c}=0.001$. }
		\label{OmegaS2dec}
	\end{figure}
	
	Figure \ref{OmegaS2} shows that when the damping is ignorable, the linear susceptibility may become negative and for definite values of $ \Omega_{s} $ or $ \omega $ its sign can be reversed. These definite values may be detected by the QFI and HSS computed with respect to the frequency of the strong laser. 
	\par
	The negative sign  of the susceptibility, implying $ n_{0} <1$,  can lead to striking interesting phenomena, such as \textit{superluminality} and \textit{parelectricity} \cite{chiao1994superluminality}.
	The fact that the linear refractive index may be less than unity implies that the phase velocity
	can be greater than the vacuum speed of light $ c $. Nevertheless, it has been proven that this phenomenon may happen without any violation of special relativity \cite{chiao1994superluminality}. In fact, the phase velocity, characterizing the velocity of the zero-crossings of the carrier wave, describes the motion of a pattern carrying no information with it \cite{chiao1997vi}.
	Moreover,  the existence of a parelectric medium indicates the possibility of stable electrostatic configurations of charges placed inside an evacuated cavity surrounded by this medium 
	as well as the levitation of an electrical charge in the vacuum above this medium \cite{chiao1995superluminality}.

	\par
	Comparing $\chi^{(1)} $ with $ HSS_{\omega_{s}} (F_{\omega_{s}} ) $
	behaviour versus $\Omega_{s} $, we see when the detuning $|\delta |$=$|\omega_{4} $-$\omega _{da}$$|$ in which $\omega_{4} $=2$\omega $+$\omega_{s} $, is high enough, the maximum point of the $ HSS_{\omega_{s}} (F_{\omega_{s}} ) $
	coincides with the point at which the sign of  $\chi^{(1)} $  is reversed. Otherwise, we see that $\chi^{(1)} $ and 
	$ HSS_{\omega_{s}} (F_{\omega_{s}} ) $
	show similar qualitative behaviour with respect to $\Omega_{s} $ (see Fig. \ref{OmegaS2a}). Furthermore, comparing their behavior versus $\omega $, we observe that either minimum or maximum point of $ HSS_{\omega_{s}} (F_{\omega_{s}} ) $ can detect the sign reversal of $\chi^{(1)} $ (see Figs. \ref{OmegaS2b} and \ref{OmegaS2c}).
	Accordingly, the HSS and QFI can be used to detect the passage of the system from subluminality into superluminality.

	\par 
	In the presence of damping, $\chi ^{(1)}$ is not real and its imaginary part, 
	characterizing the linear absorption, can be related to the QFI and HSS.
	Analyzing the variations of $\text{Im}[\chi ^{(1)}]$ and  $ HSS_{\omega_{s}} (F_{\omega_{s}} ) $
	versus $\Omega_{s}$ in Fig. \ref{OmegaS2deca}, we find that when 
	$\gamma_{d} >> \gamma _{c}$ as well as
	$\gamma_{c} $ is small, and 
	the detuning $|\delta |$=$|\omega_{4} $-$\omega_{da} $$|$, in which $\omega_{4} $=2$\omega $+$\omega_{s} $, is high enough,  the maximum point of  $\text{Im}[\chi ^{(1)}]$ and $ HSS_{\omega_{s}} (F_{\omega_{s}} ) $  coincide.  Moreover, under the aforementioned conditions,  these measures increase or decrease with each other. Now assume that  the detuning $|\delta |$=$|\omega_{4} $-$\omega_{da} $$|$ is not high enough.  Under these conditions,  $\text{Im}[\chi ^{(1)}]$ and  $ HSS_{\omega_{s}} (F_{\omega_{s}} ) $
	exhibit similar qualitative behavior with respect to $\Omega_{s} $ such that, as shown in Fig. \ref{OmegaS2decb}, they monotonously decrease with an increase in  $\Omega_{s}  $.

	In Fig. \ref{OmegaS2decc}, we can compare the behavior of  $\text{Im}[\chi ^{(1)}]$ with $ HSS_{\omega_{s}} (F_{\omega_{s}} ) $ versus $ \omega $. It is found that
	$\gamma_{d} >> \gamma _{c}$ as well as
	$\gamma_{c} $ is sufficiently small, both $\text{Im}[\chi ^{(1)}]$ and  $ HSS_{\omega_{s}} (F_{\omega_{s}} ) $
	are minimized  for the value of $\omega $ at which the detuning $\delta $=$\omega_{4} $-$\omega_{da} $ becomes zero. In addition, they decrease and increase with each other, i.e., there is a harmonious relationship between their variations. 
	
	\par 
	Furthermore, comparing $\text{Im}[\chi ^{(1)}]$ and  $ HSS_{\omega_{s}} (F_{\omega_{s}} ) $ behaviors with respect to $\omega_{dc} $ in Fig. \ref{OmegaS2decd}, we observe that when $\delta $=0,   both measures are minimized  for $\omega_{dc} $=$\omega_{s} $, leading to $\Delta $=$\omega_{dc} $-$\omega_{s} $ for which the EIT exhibit the best efficiency. 
	
	\par These analyses show that the HSS can be used to monitor the linear absorption of the weak wave in the medium. In addition, this monitoring can be applied to improve the frequency estimation.

	\subsubsection{Quantum statistical speeds with respect to $ \omega $ }\label{RelationOmega2}

	\begin{figure}[ht]
		\subfigure[] {\includegraphics[width=7cm]{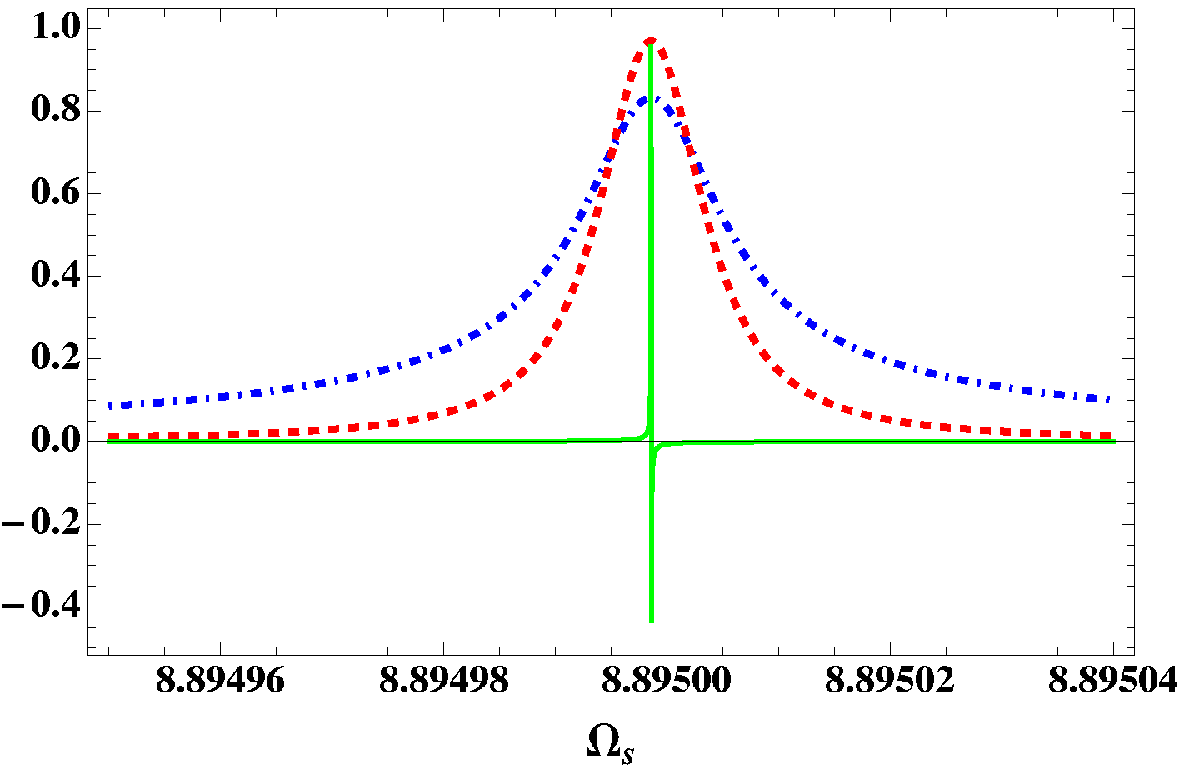}\label{Omega2a}} 
		\hspace{0.5mm}
		\subfigure[] { \includegraphics[width=7cm]{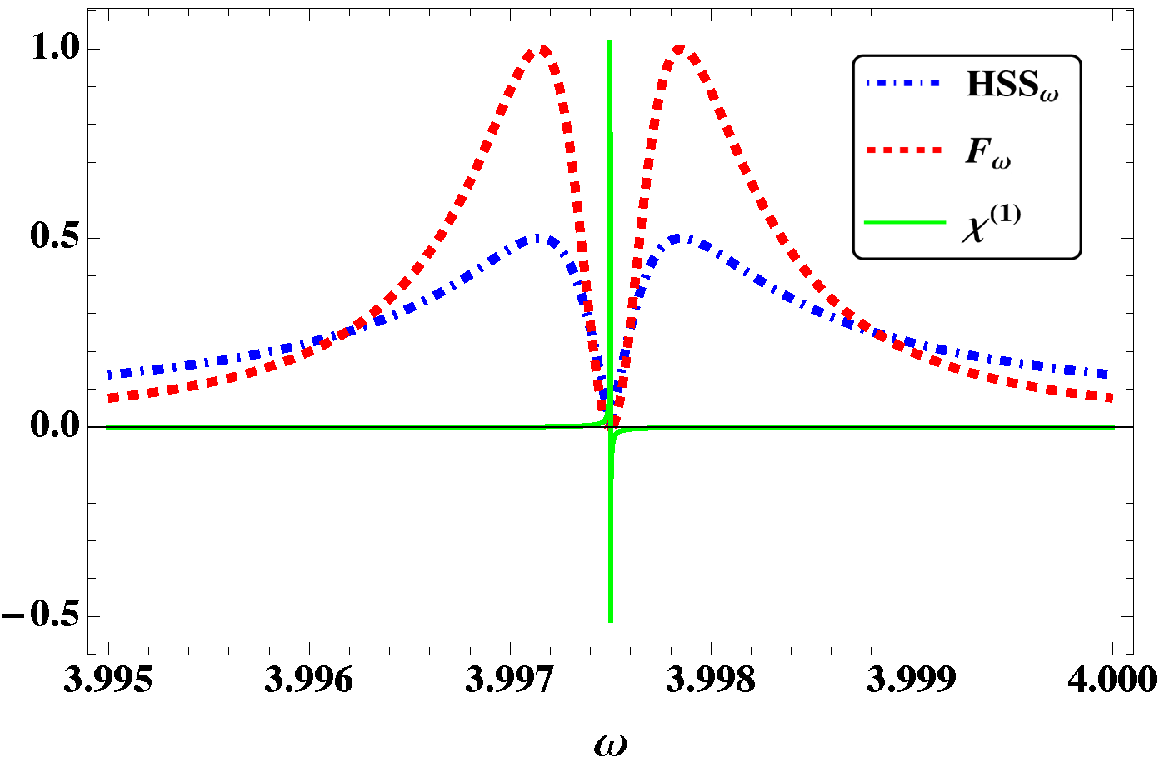}\label{Omega2b}}
		\hspace{0.5mm}
		\subfigure[] { \includegraphics[width=7cm]{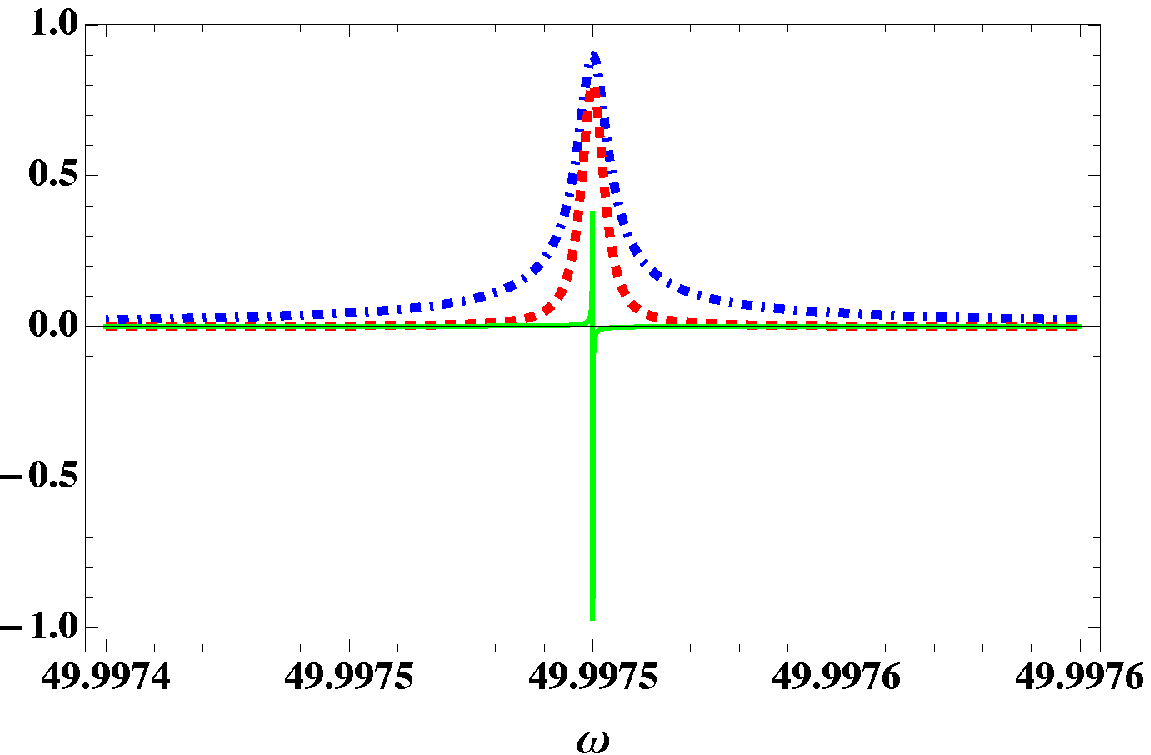}\label{Omega2c}}
		\caption{Three-level system in the absence of damping: (a) Normalized quantum Fisher information $ F_{\omega} $, Hilbert-Schmidt speed $ HSS_{\omega} $, and  linear susceptibility  $ \chi^{(1)} $ as functions of $ \Omega_{s} $  for 
			$\Omega=0.00001 ,  \omega_{da}=20, \omega=4.65, \omega_{dc}=1.8, \omega_{s}=1.81$; 
			(b)  The same quantities versus $ \omega $  for 
			$\Omega=0.001 ,  \Omega_{s}=10, \omega_{da}=20, \omega_{dc}=2, \omega_{s}=2.01$. (c)  The same quantities versus $ \omega $  for 
			$\Omega=0.00001 ,  \Omega_{s}=100, \omega_{da}=210, \omega_{dc}=10, \omega_{s}=10.01$.}
		
		\label{Omega2}
	\end{figure}

	\begin{figure}[ht]
		\subfigure[] {\includegraphics[width=7cm]{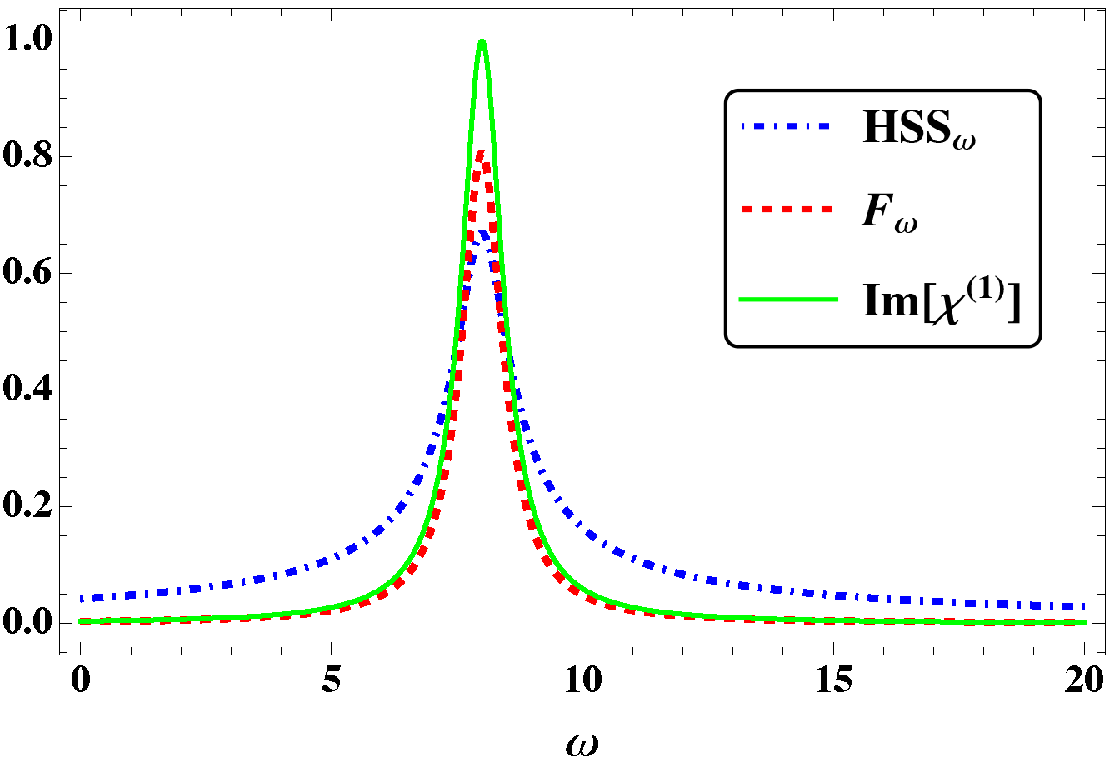}\label{Omega2deca}} 
		\hspace{0.5mm}
		\subfigure[] { \includegraphics[width=7cm]{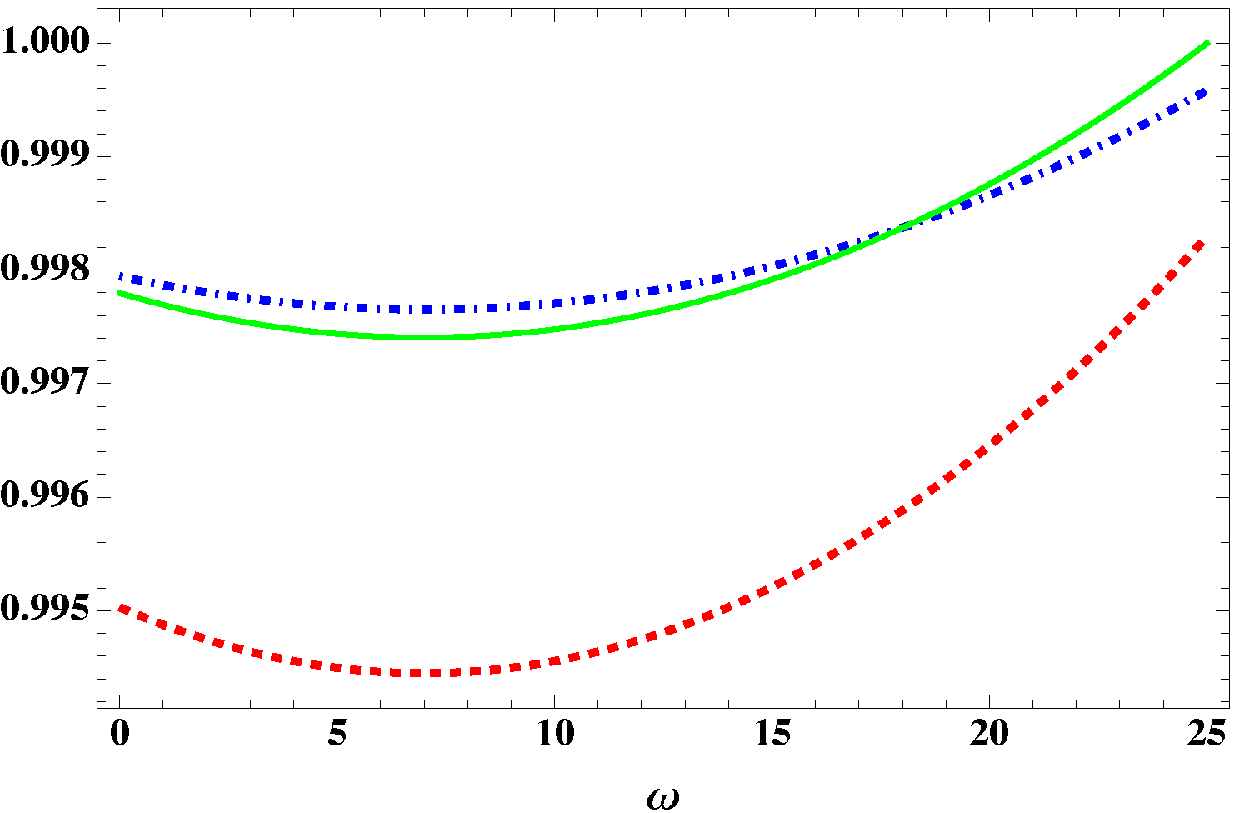}\label{Omega2decb}}
		\hspace{0.5mm}
		\subfigure[] { \includegraphics[width=7cm]{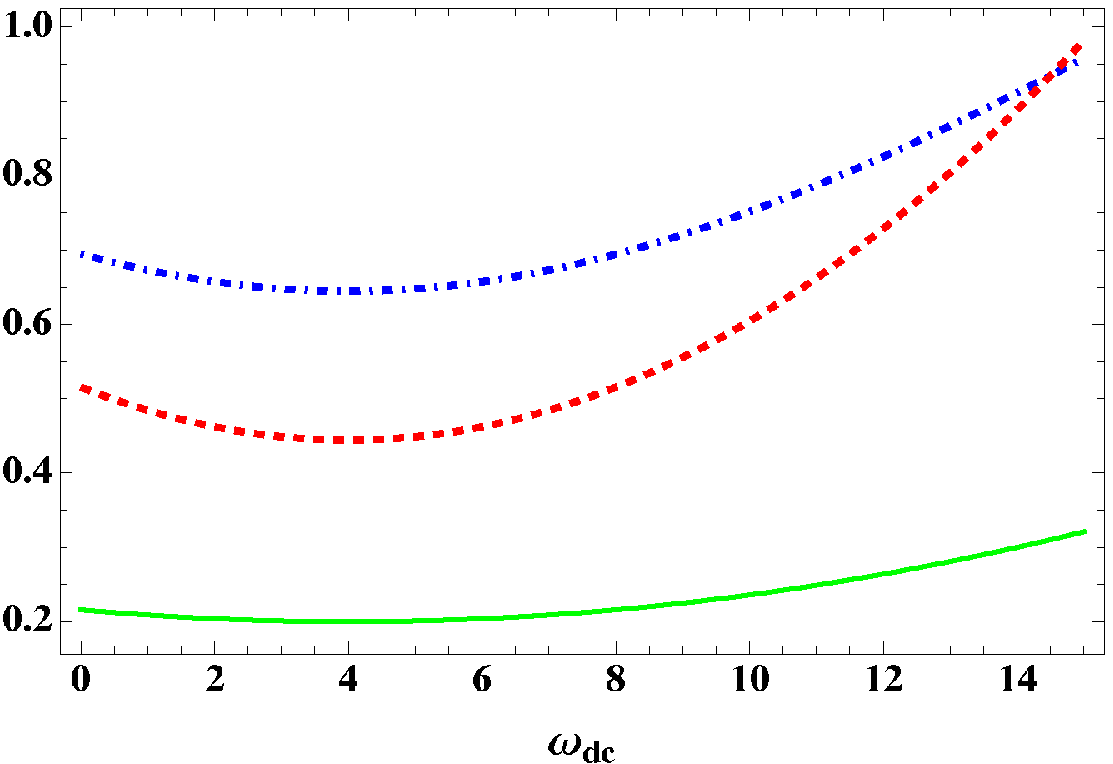}\label{Omega2decc}}
		\caption{Three-level system in the presence of damping: (a) Comparison among normalized quantum Fisher information $ F_{\omega} $, Hilbert-Schmidt speed $ HSS_{\omega} $, and the   imaginary part of the linear susceptibility  $ \text{Im}[\chi^{(1)}] $  versus $ \omega $  for 
			$\Omega= 0.00001, \Omega_{s}= 10$, $  \omega_{da}=20,  \omega_{dc}=4, \omega_{s}=4.001,  \gamma_{c}=100,  \gamma_{d}=0.001$; 
			(b)  The same quantities versus $ \omega $  for 
			$\Omega_{s}= 1000$, $  \omega_{da}=18,  \gamma_{c}=100,  \gamma_{d}=1$  (c) The same quantities versus  $ \omega_{dc} $  for 
			$   \omega=8, \omega_{s}=4,  \gamma_{c}=0.02,  \gamma_{d}=0.01$. }
		\label{Omega2dec}
	\end{figure}
	
	In this section, we investigate the usefulness of HSS, computed with respect to $ \omega $, for detecting  parelectricity, analyzing the absorption experienced by the weak wave in the medium, and improving its frequency estimation. 
	\par 
	Computing the HSS and QFI with respect to $ \omega $ we again find that they can detect the sign-reversal of the linear susceptibility.  The behavior of  $\chi^{(1)} $ and $ HSS_{\omega} (F_{\omega} ) $
	versus $\Omega_{s} $ and $\omega$ is compared in Fig. \ref{Omega2} plotted when the damping is ignorable. Figure \ref{Omega2a} shows that when the detuning $|\delta |$, is high enough, the maximum point of the $ HSS_{\omega} (F_{\omega} ) $
	coincides with the point at which   $\chi^{(1)} $  experiences the sign reversal, similar to the behavior observed in Fig. \ref{OmegaS2a}. 
	Moreover, the sign reversal of $\chi^{(1)} $ can also be detected by searching for the minimum or maximum point of $ HSS_{\omega} (F_{\omega} ) $ versus $\omega $, as shown in Figs. \ref{Omega2b} and \ref{Omega2c}. It should be noted that although maximum and minimum points of $ F_{\omega_{s}} $ as well as $ F_{\omega} $ are witnesses of the sign reversal occurring for  $\chi^{(1)}$, we cannot generally conclude that their qualitative behaviors are completely similar, because they may vary oppositely. In other words, for example, when $ F_{\omega_{s}} $ is maximized  (minimized) with respect to $ \Omega_{s} $ or $ \omega $, we see that $ F_{\omega} $ may be both maximized and minimized (compare Figs. \ref{OmegaS2} with  Fig.  \ref{Omega2}).
	
	\par 
	According to the aforementioned analyses, we find that not only the HSS and QFI, computed with respect to the frequencies of the driving waves, are efficient tools to detect the passage of the system from subluminality into superluminality, but also monitoring the variations of the linear refractive index can be applied to enhance the frequency estimation.

	\par 
	Figures \ref{Omega2deca}  and  \ref{Omega2decb} illustrate  the variations of $\text{Im}[\chi ^{(1)}]$ and  $ HSS_{\omega} (F_{\omega} ) $ versus $\omega$ in the presence of damping. We see that when
	$\gamma_{c} \geqq \gamma _{d}$  both $\text{Im}[\chi ^{(1)}]$ and  $ HSS_{\omega} (F_{\omega} ) $ exhibit harmonious behavior such that they 
	are either maximized or minimized  for the value of $\omega $ at which the detuning $\delta $ becomes zero.
	\par 
	In addition, investigating the behavior of measures with respect to $\omega_{dc} $, we find  that when $\delta =0$ and $\gamma_{c} \geqq \gamma _{d}$,   $\text{Im}[\chi ^{(1)}]$ and  $ HSS_{\omega} (F_{\omega} ) $  are minimized  for $\omega_{dc} $=$\omega_{s} $,  leading to the best efficiency in the EIT process (see Fig. \ref{Omega2decc}). 
	
	\par
	Therefore, in the presence of the damping, there are close relationships among the (HSS, QFI), calculated with respect to the frequency of the driving waves, and the linear absorption experienced by the weak wave.

	\section{Conclusions}\label{Conclusion}
	In this paper, we have investigated the application of the quantum Fisher information (QFI) and Hilbert-Schmidt speed (HSS) in characterizing the linear and nonlinear responses of the medium, consisting of the studied three and four-level atoms, to applied optical fields in the sum-frequency generation and electromagnetically induced transparency (EIT) processes. Because the QFI and HSS are computed focusing on one of the atoms located in the atomic medium, we can predict the collective optical behavior of the atomic ensemble by analyzing the quantum information extracted from one of its elements. This is one of the most important
	results of the paper.

	\par
	
	Moreover, we found that by monitoring the variations of the (linear and nonlinear) refractive indices of the medium, we can determine the optimal strategy in the process of frequency estimation. This result is of great practical importance. In particular, it is well known that the \textit{Z-scan technique }  \cite{sheik1990sensitive,desalvo1992self,singh2015z} can be applied to measure intensity-dependent nonlinear susceptibilities of nonlinear optical materials.  Therefore,  the nonlinear refractive index $ n_{2} $ can be determined and monitored experimentally to achieve the best accuracy in the process of frequency estimation.

	\par
	
	We also observed that the regions at which the linear susceptibility is negative and hence the linear refractive index is smaller than $ 1 $,  are detectable by the HSS and QFI. Provided that the atomic density $ N $ is sufficiently large, one may expect that in a medium whose linear susceptibility is negative, we have $ \epsilon^{(1)} =1+ \chi^{(1)}<0   $. Therefore, negative permittivity can also be revealed by the technique presented in this paper. Accordingly, the HSS and QFI are introduced as potential candidates to characterize ferrite materials with negative permeability as well as metamaterials in which permittivity ($\epsilon$), and magnetic permeability ($ \mu $) are simultaneously negative. Our work motivates new research on the practical application of the HSS and QFI in designing new materials with artificial magnetism, negative refractive index, and negative refraction, such as super-lenses \cite{smith2004metamaterials,ramakrishna2005physics,liu2011metamaterials}.

	\par
	
	One of the most important motivations for applying nonlinear effects and especially harmonic generation in microscopy is providing enhanced longitudinal and transverse resolution \cite{boyd2020nonlinear}.  Nonlinear effects are excited optimally in the region of maximum intensity of a focused laser beam, and hence resolution can be improved. Moreover,  the advantage that the signal is far removed in frequency from unwanted background light, resulting from linear scattering of the incident laser beam, is also offered by microscopy based on harmonic generation.
	Therefore, our results propose the application of HSS and QFI for enhancing image resolutions which is of great importance in various fields of science and technology \cite{muller19983d,fiderer2021general}.
	
	\par
	
	Optically induced damage of optical components is one of the most important topics from the practical point of view. Optical damage \cite{ammosov1986tunnel,bloembergen1974laser,kong2020recent} is significant as it can finally restrict the maximum amount of power transmittable through a particular optical material. Therefore, optical damage limits the efficiency of many nonlinear optical processes by restricting the maximum field strength $ E $ that can be applied to excite the nonlinear response without rising optical damage. In this paper, we have shown that the HSS and QFI can determine the  Rabi frequency, explicitly related to the strength of the applied field, which excites the optimal nonlinear response. In this context,  the HSS and QFI are proposed to detect optical damage and exceed the damage thresholds in various materials.

	\section*{Declaration of competing interest}
	We have no competing interests.

	\bibliographystyle{apsrev4-2} 
	\bibliography{apssamp2} 
	
\end{document}